\documentclass[intlimits,twoside,a4paper]{article}
\usepackage{amsmath,amssymb}
\usepackage{graphicx}
\usepackage[T2A]{fontenc}
\usepackage[cp1251]{inputenc}
\usepackage{siunitx}

\usepackage{cmpj2}

\usepackage{color}

\articletype{Proceedings Paper}

\issue{2013}{16}{3}{31706}
\doinumber{10.5488/CMP.16.31706}

\title[Dielectric properties of PLZT-$x/65/35$ ($2\leqslant x \leqslant 13$)]
{Dielectric properties of PLZT-$x/65/35$ ($2\leqslant x \leqslant 13$) under mechanical stress, electric field and temperature loading}

\author[K. Pytel \textsl{et al.}]{K. Pytel\refaddr{ad1}, J. Suchanicz\refaddr{ad2}\thanks{E-mail: sfsuchan@up.krakow.pl}\ ,
M. Livinsh\refaddr{ad3}, A. Sternberg\refaddr{ad3}}

\addresses{
\addr{ad1}Institute of Technology, Pedagogical University, 2 Podchorazych St., 30--084 Krakow, Poland
\addr{ad2}Institute of Physics, Pedagogical University, 2 Podchorazych St., 30--084 Krakow, Poland
\addr{ad3}Institute of Solid State Physics, University of Latvia, 8 Kengaraga St., LV--1063 Riga, Latvia
}

\date{Received  October 23, 2012}
\authorcopyright{K. Pytel, J. Suchanicz, M. Livinsh, A. Sternberg, 2013}

\begin{document}

\maketitle

\begin{abstract}

We investigated the effect of uniaxial pressure ($0 \div 1000$~bars) applied parallely to the ac electric field on dielectric properties of PLZT-$x/65/35$ ($2 \leqslant x \leqslant 13$) ceramics. There was revealed a significant effect of the external stress on these properties. The application of uniaxial pressure leads to the change of the peak intensity of the electric permittivity ($\varepsilon$), of the frequency dispersion as well as of the dielectric hysteresis. The peak intensity $\varepsilon$  becomes diffused/sharpened and shifts to a higher/lower temperatures with increasing the pressure. It was concluded that the application of uniaxial pressure induces similar effects as increasing the Ti ion concentration in PZT system. We interpreted our results based on the domain switching processes under the action of combined electromechanical loading.
\pacs 77.84.Lf, 77.80.bg
\end{abstract}

\section{Introduction}

Single crystals, ceramics, polymers and ceramic-polymer composites which show ferroelectric behavior are used in many applications in electronics and optics. A large number of applications of ferroelectrics exploit dielectric, piezoelectric, pyroelectric and electro-optic properties. Ferroelectric crystals are commonly used in many areas of science and technology. Possibilities of the practical use of these ceramics are associated with  their outstanding electro-mechanical properties. A very small amount of dopants may be sufficient to have a noticeable effect on macroscopic properties of these technologically important materials. Ferroelectrics can be modified not only by purposeful doping but also by impurities or by defects formed at high processing temperatures. Softening and hardening of ferroelectrics by acceptor and donor doping are the key tools to control their electrical and mechanical properties. These properties include e.g. switching and hysteresis behavior, frequency dispersion and nonlinearity of electromechanical properties. PZT ceramics can be doped with ions to form soft and hard lead zirconate titanates.
The soft one have a higher permittivity, larger electrical losses, higher piezoelectric coefficient and are easy to polarize and depolarize. They can be used for applications requiring very high piezoelectric properties. The hard one have a lower permittivity, smaller electrical losses, lower piezoelectric coefficients and are more difficult to polarize and depolarize. They can be used for rugged applications \cite{1}.
Ferroelectrics could be modified by substitutions of Pb$^{2+}$ ions in PZT ceramics by La$^{2+}$ one (i.e. PLZT). For PLZT-$x$/65/35 with $x \leqslant 6$, a paraelectric to ferroelectric phase transition occurs as the temperature is lowered through the Curie point $T_\textrm{c}$ and a second structural transition occurs at a lower temperature $T_\textrm{t}$ \cite{2}. For PLZT-$x$/65/35 with $x\geqslant7$ no paraelectric to ferroelectric transition occurs as the temperature is lowered through $T_\mathrm{c}$ and ceramic shows a broad peak in the low frequency dielectric constant, $\epsilon(T)$  \cite{3}. Compositions $x$/65/35 with $x>6$ exhibits typical relaxor behaviors such as frequency-dependent electric permittivity maximum, existence of polar nanoregions at temperatures above the diffuse phase transition point and broad distribution of relaxation times  \cite{4}. The incorporation of aliovalent lanthanum into the lattice enhances the densification rates of the PZT ceramics, leading to pore-free homogeneous microstructures.
In many device applications, ceramics are subjected to combined electric and stress fields, so a better understanding of phenomena occurring under electromechanical loading is essential. In the present study the results of dielectric properties measurements of PLZT  ceramics with a Zr/Ti ratio of 65/35 and variable La content (2--13 at.\%) under uniaxial pressure applied parallel to the ac electric field were presented.

\section{Experimental procedure}
\subsection{Material}

PLZT-$x/65/35$ with a different content of lanthanum powders were acquired by two-stage co-pre\-ci\-pi\-ta\-tion method from mixed solution of inorganic salts ZrOCl$_{2}\cdot8$H$_{2}$O, TiCl$_{4}$, La(NO$_{2}$)$_3\cdot6$H$_{2}$O and Pb(NO$_{3}$)$_{2}$ well described in paper \cite{5}. At the first stage, the hydroxypolymer of TiO$_{2}$-ZrO$_{2}$-La$_{2}$O$_{3}$ is obtained by co-precipitation with 25\% NH$_{4}$OH from a mixed solution of the corresponding metallic salts. At the second stage, PbO was introduced into the mixture of TiO$_{2}$-ZrO$_{2}$-La$_{2}$O$_{3}$. Ceramic samples were prepared by a two-stage hot-pressing technology \cite{5}. The first stage was performed at $1150 \div 1180$\SI{}{\degreeCelsius} for $1$ hour in vacuum at pressure 200~MPa. The second stage was performed at $1150 \div 1200$\SI{}{\degreeCelsius} for $1 \div 40$ hours depending on the size ($15 \div 90$~mm diameter) at pressure 200~MPa in the air or in O$_{2}$ rich atmosphere. As a result, high density and transparent ceramics were obtained.

\begin{figure}[!b]
\centerline{
\includegraphics[width=0.4\textwidth]{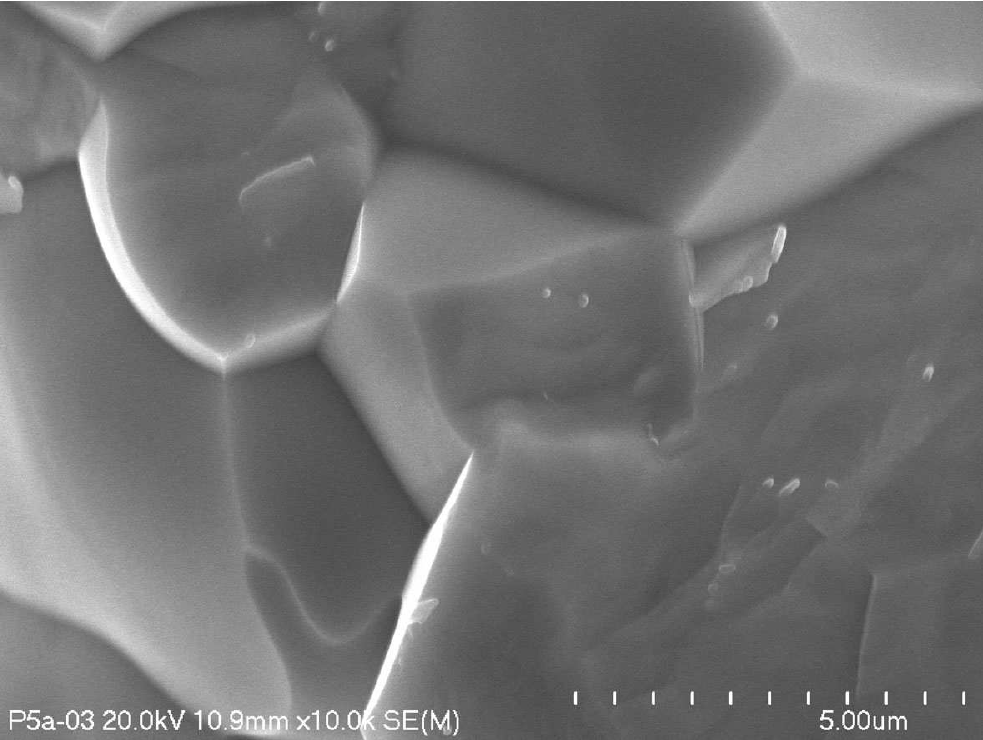}
\hspace{1mm}
\includegraphics[width=0.4\textwidth]{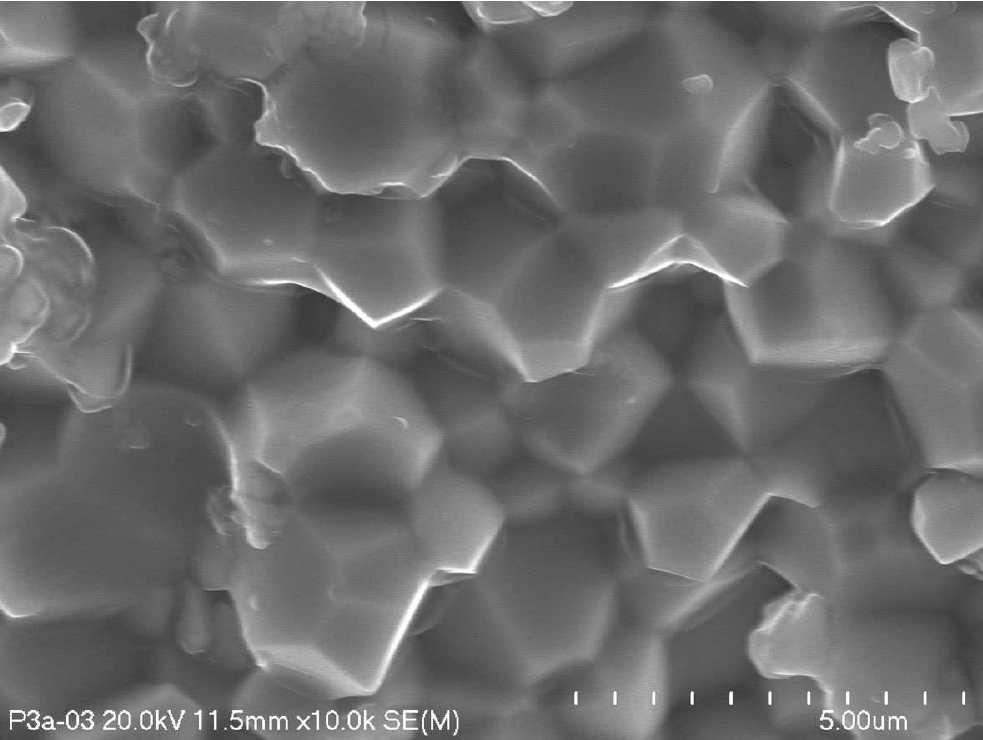}
}
\centerline{(a) \hspace{0.39\textwidth} (b)}
\vspace{1mm}
\centerline{
\includegraphics[width=0.4\textwidth]{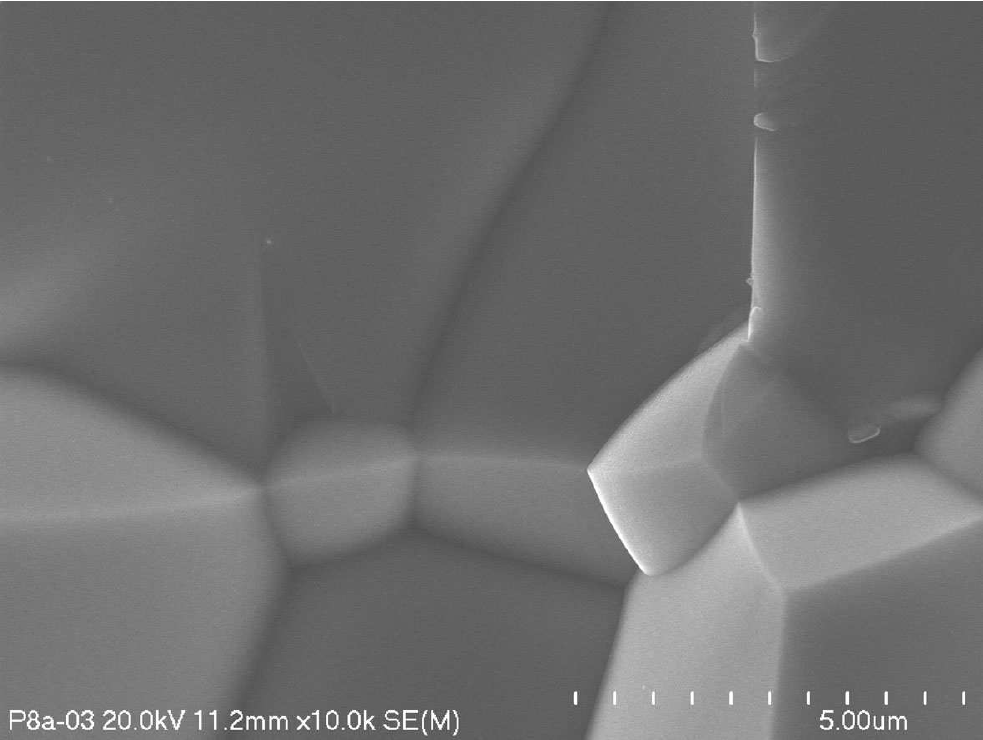}
\hspace{1mm}
\includegraphics[width=0.4\textwidth]{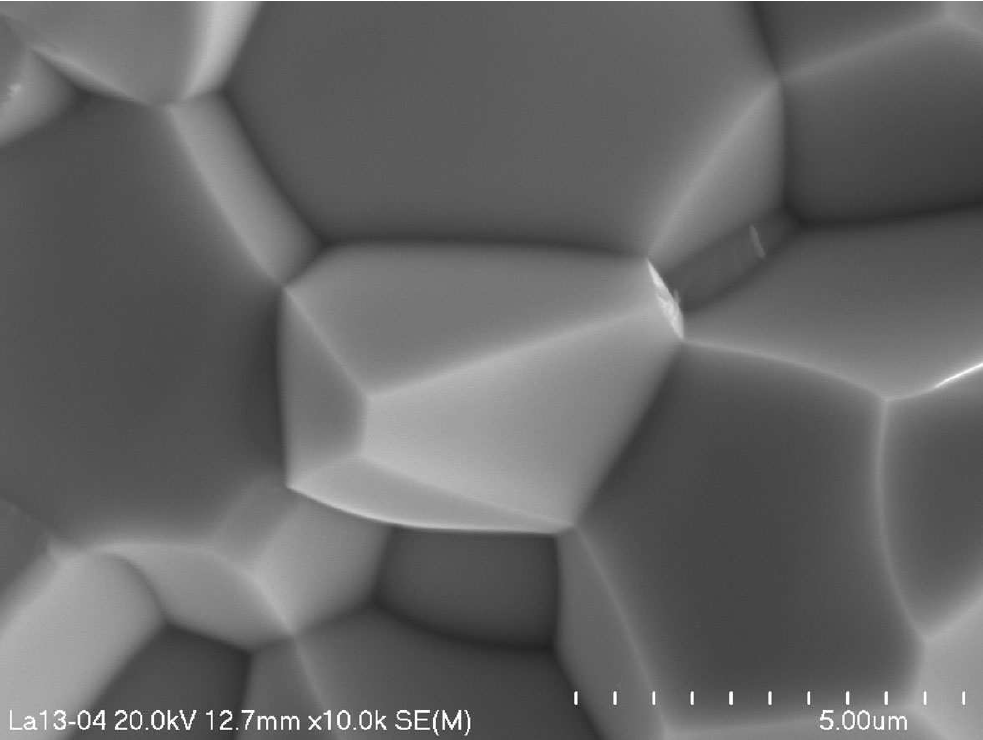}
}
\centerline{(c) \hspace{0.39\textwidth} (d)}
\caption{SEM image of the fracture surface of PLZT-$x/65/35$ ceramics ($x=2\,, 6,\, 10$ and $13$).} \label{fig1}
\end{figure}

\subsection{Microstructure measurements}

The scanning electron microscope (SEM) Hitachi S-4700, equipped with an energy dispersive X-ray spectrometer (EDS) having Si(Li) X-ray detector, was used for the investigation of the microstructure of ceramics. The EDS analysis was performed using the Noran-Vantage system.

\subsection{Dielectric measurements}

The dielectric measurements were carried out in a weak electric field (30 V/cm) using BM 595A LCR meter in a temperature range from room temperature to 460\SI{}{\degreeCelsius}. The samples were previously electroded by silver paste. The temperature of samples was controlled by a thermocouple with the accuracy of $\pm$0.1\SI{}{\degreeCelsius}. Prior to the experiments, the samples were heated for at least 1h at about 500\SI{}{\degreeCelsius} to release both internal strains and those at the electrode/sample interface. The temperature dependence of the permittivity was measured during heating and cooling processes at the rate of l00\SI{}{\degreeCelsius}/h. Compressive stress within the range of ($0 \div 1000$) bar was applied parallel to the measuring electric field with the use of a lever and a weight.

\section{Results and discussion}

\begin{figure}[!tb]
\centerline{
\includegraphics[width=0.45\textwidth]{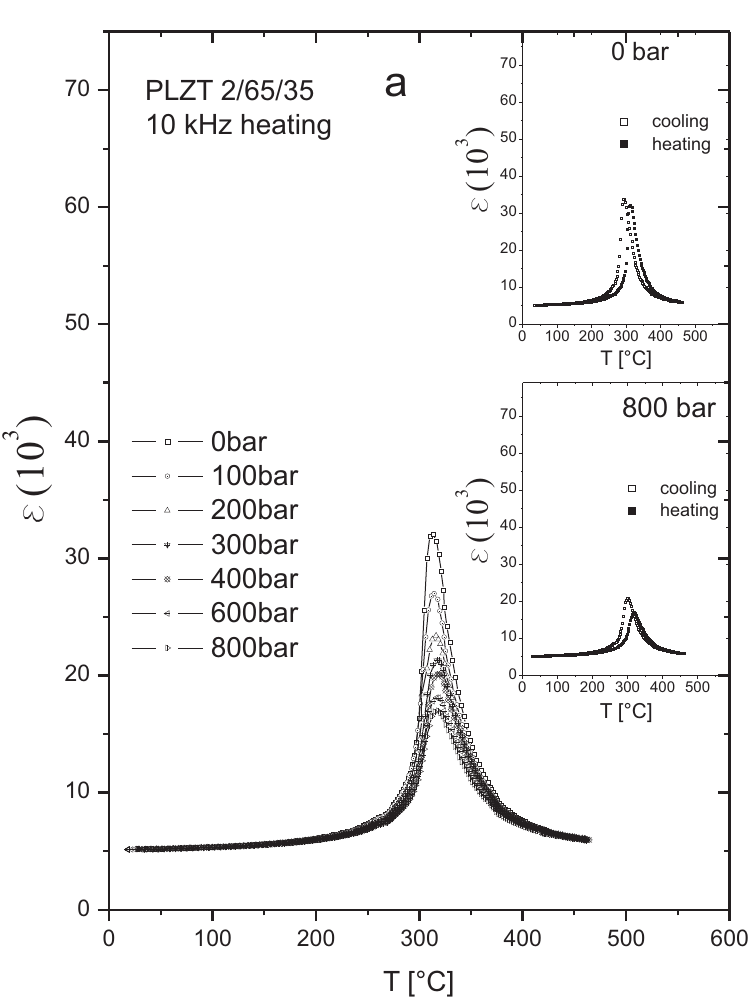}
\hspace{1mm}
\includegraphics[width=0.45\textwidth]{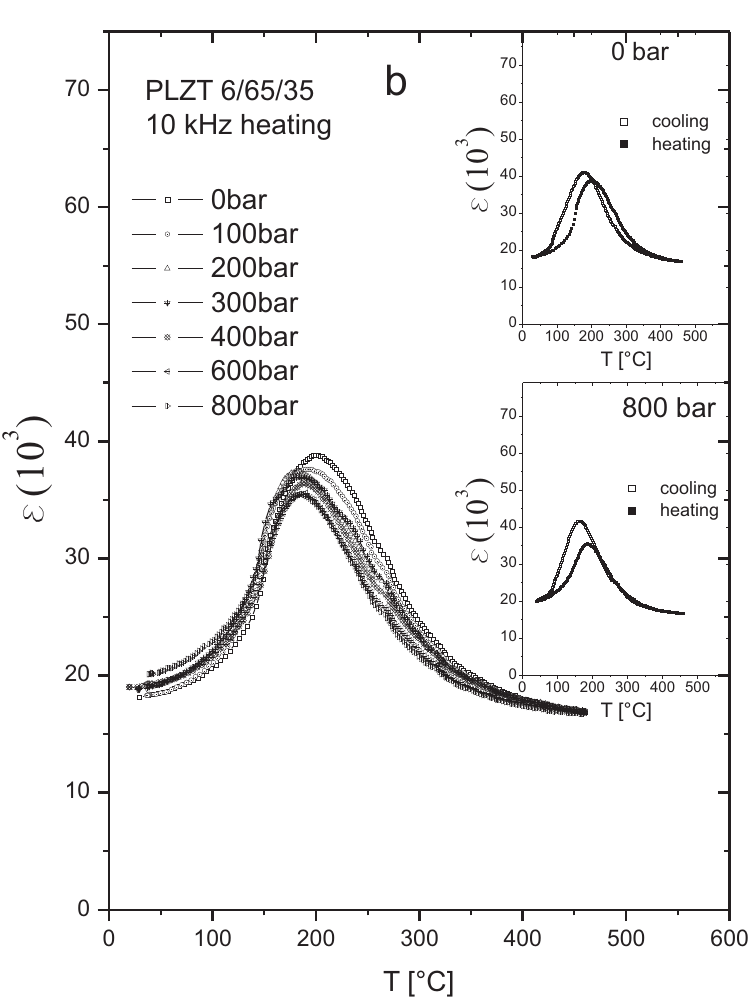}
}
\centerline{
\includegraphics[width=0.45\textwidth]{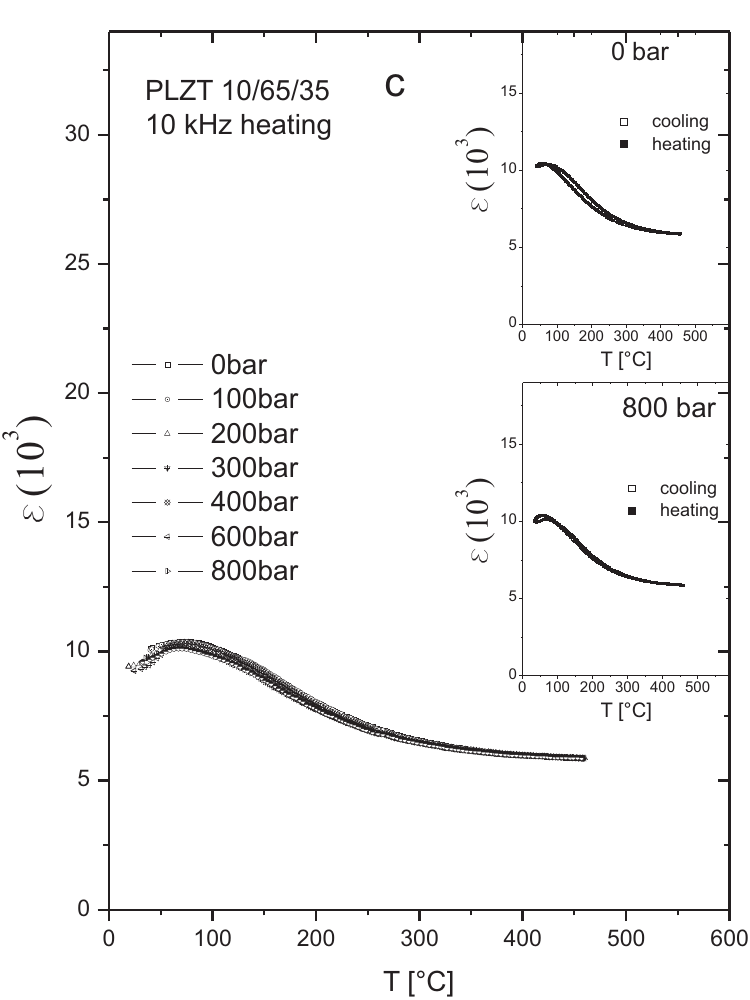}
\hspace{1mm}
\includegraphics[width=0.45\textwidth]{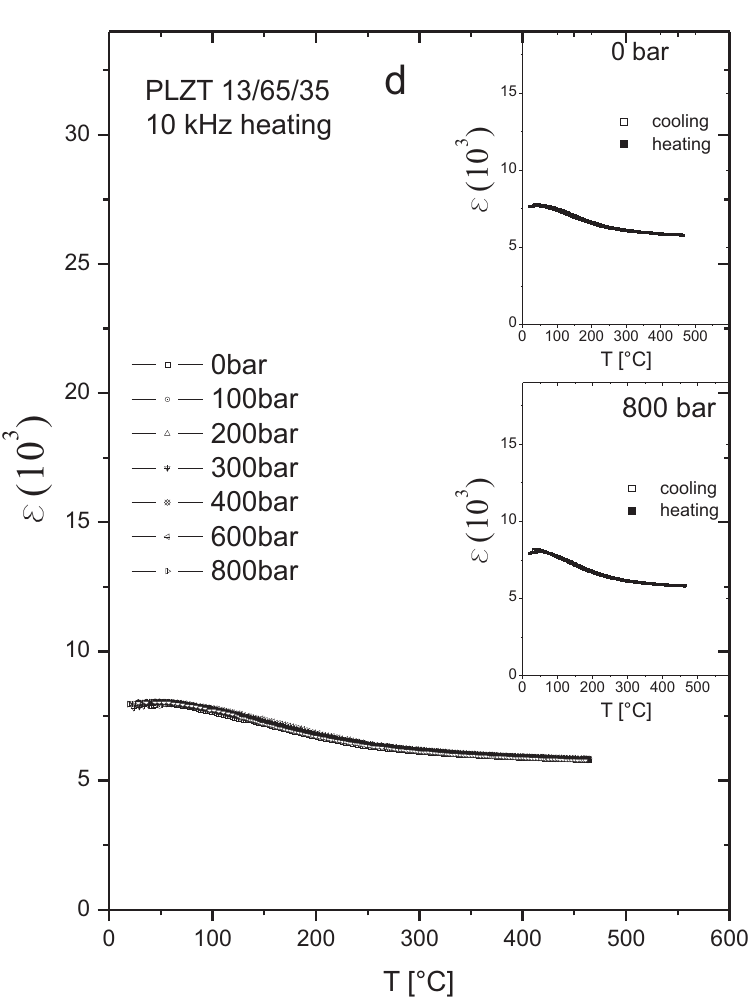}
}
\caption{Temperature/pressure dependence of the electric permittivity $\varepsilon$ of PLZT-$x/65/35$ ceramics with $x=2$ (a), $6$ (b), $10$ (c) and $13$ (d) (on heating, $f=1$~kHz). Inserts show the same on heating and cooling for $p=0$~bar and $800$~bar, respectively.} \label{fig2}
\end{figure}

Figure~\ref{fig1} shows the cross-section morphological feature of PLZT-$x/65/35$ with $x=2,\, 6,\, 10$ and $13$. As can be seen from this figure, dense and crack-free ceramics without second phases were sintered. The average grain size of the regular crystal-shape ceramics is ca. $5 \div 8$~\SI{}{\micro\metre} for PLZT-$x/65/35$ with $x=2,\, 6, 10$ and $13$. The crystalline boundaries are clearly observed. The EDS analyses completed for individual grains in different compositions indicated a homogenous distribution of all elements in ceramics. Chemical analysis of the obtained samples performed by emission microanalyzer for Pb, Zr, Ti and La agreed well with the nominal composition of PLZT.

\begin{figure}[!tb]
\centerline{
\includegraphics[width=0.45\textwidth]{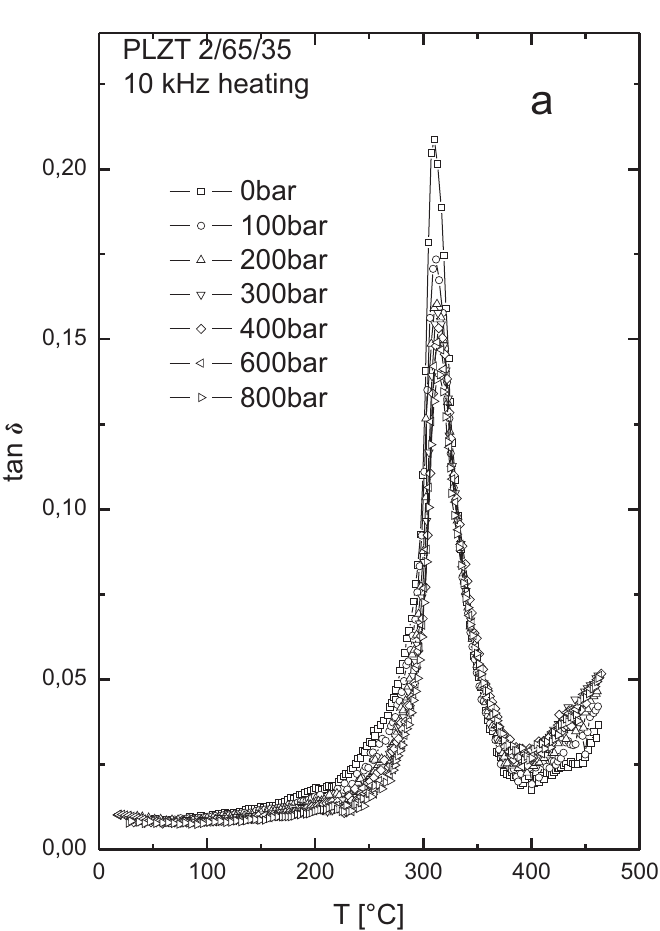}
\hspace{1mm}
\includegraphics[width=0.45\textwidth]{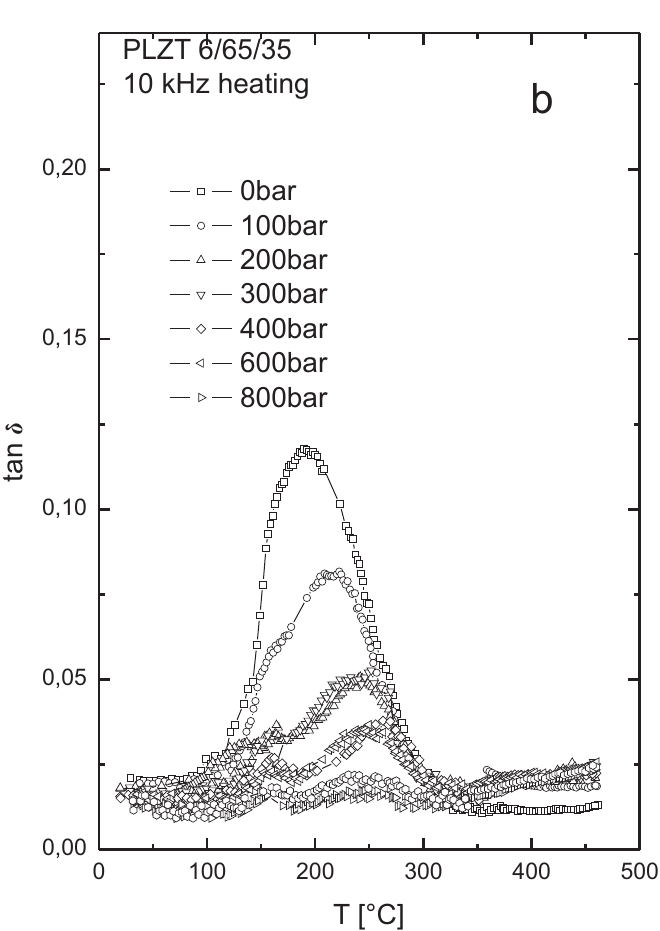}
}
\centerline{
\includegraphics[width=0.45\textwidth]{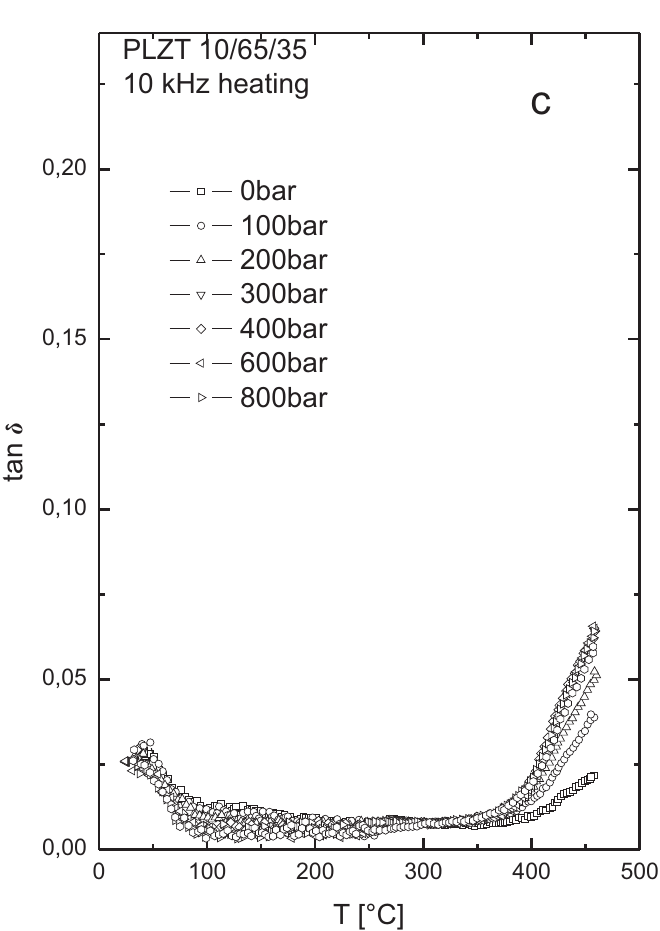}
\hspace{1mm}
\includegraphics[width=0.45\textwidth]{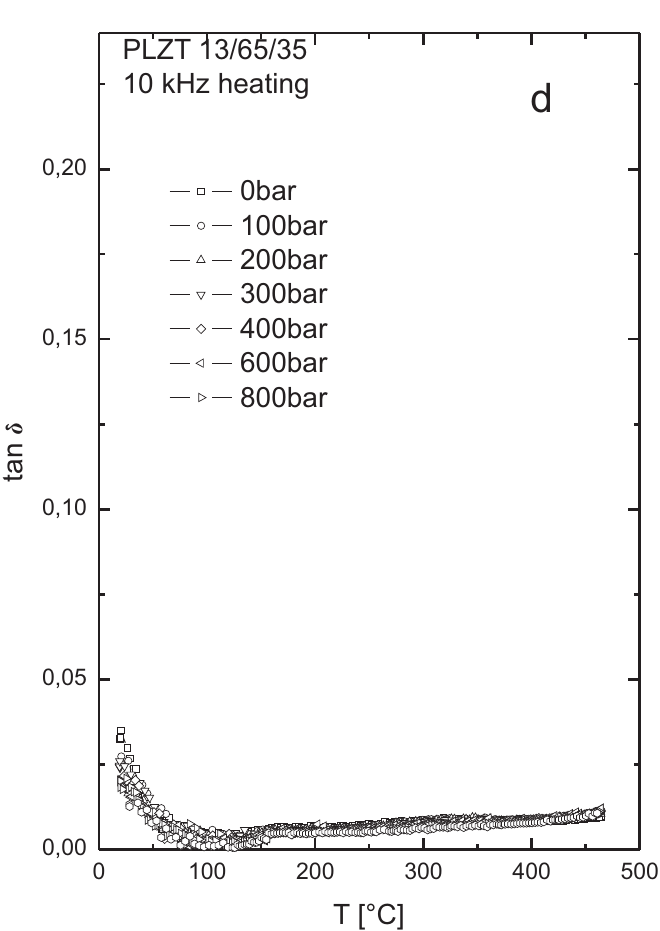}
}
\caption{Temperature/pressure dependence of the dielectric loss of PLZT-$x/65/35$ ceramics with $x=2$ (a), $6$ (b), $10$ (c) and $13$ (d) on heating ($f=10$~kHz).}
\label{fig3}
\end{figure}

\begin{figure}[!tb]
\centerline{
\includegraphics[width=0.45\textwidth]{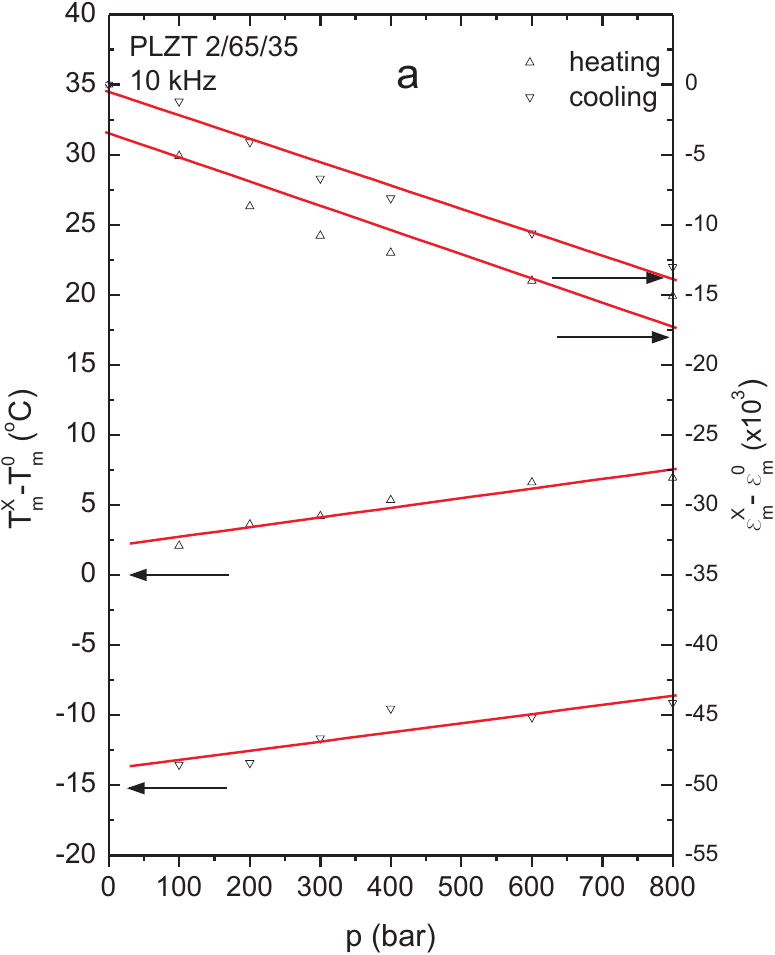}
\hspace{1mm}
\includegraphics[width=0.45\textwidth]{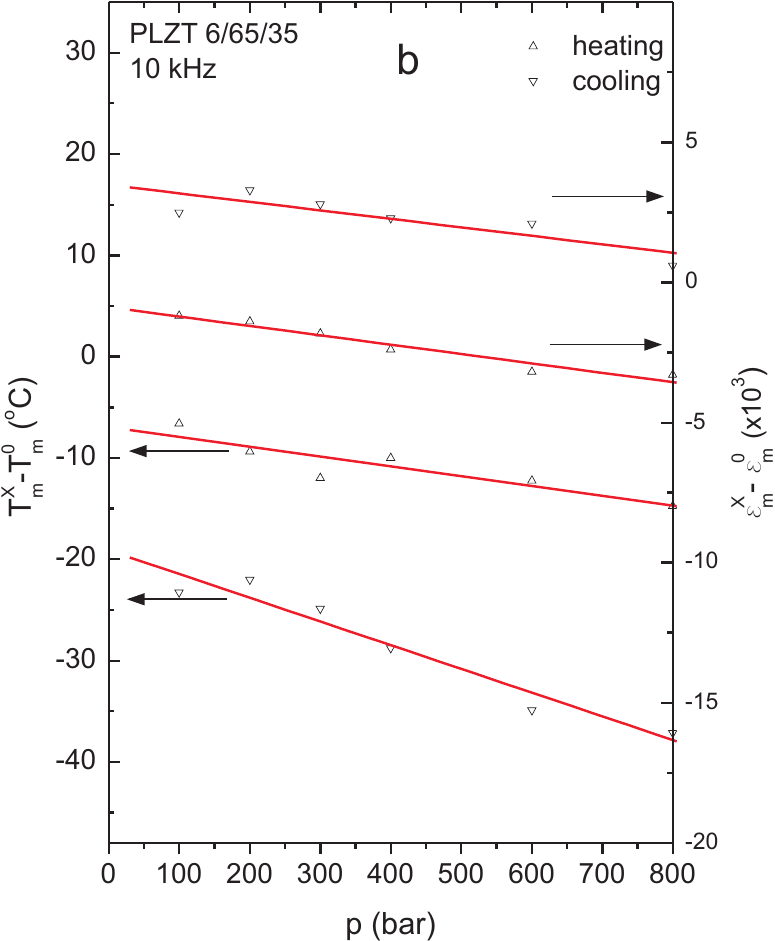}
}
\centerline{
\includegraphics[width=0.45\textwidth]{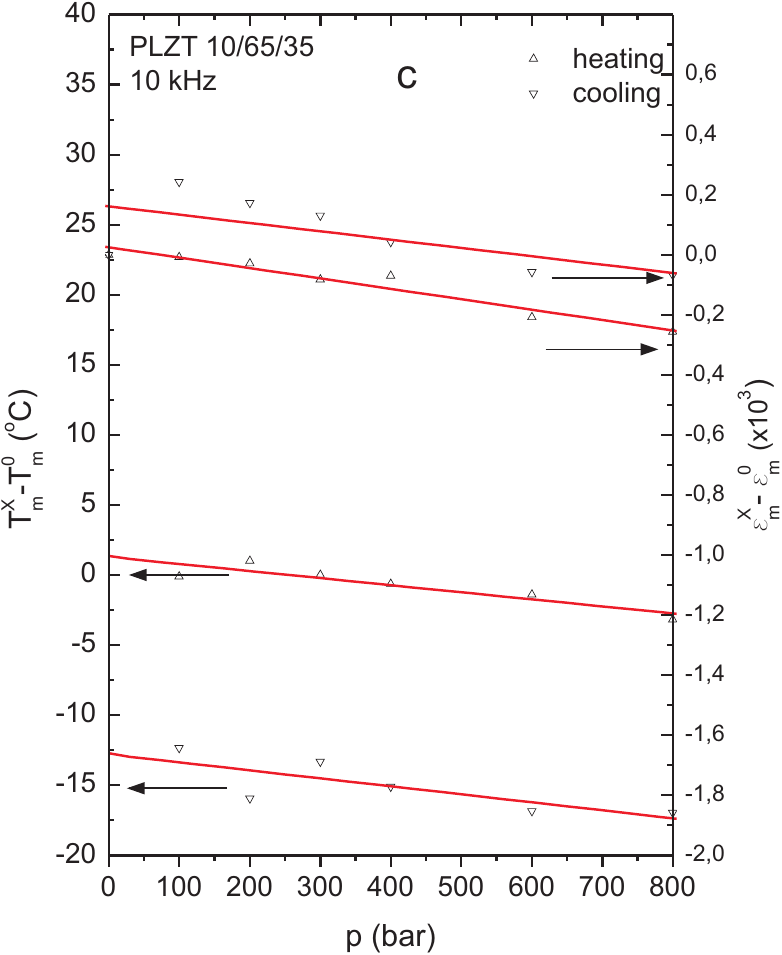}
\hspace{1mm}
\includegraphics[width=0.45\textwidth]{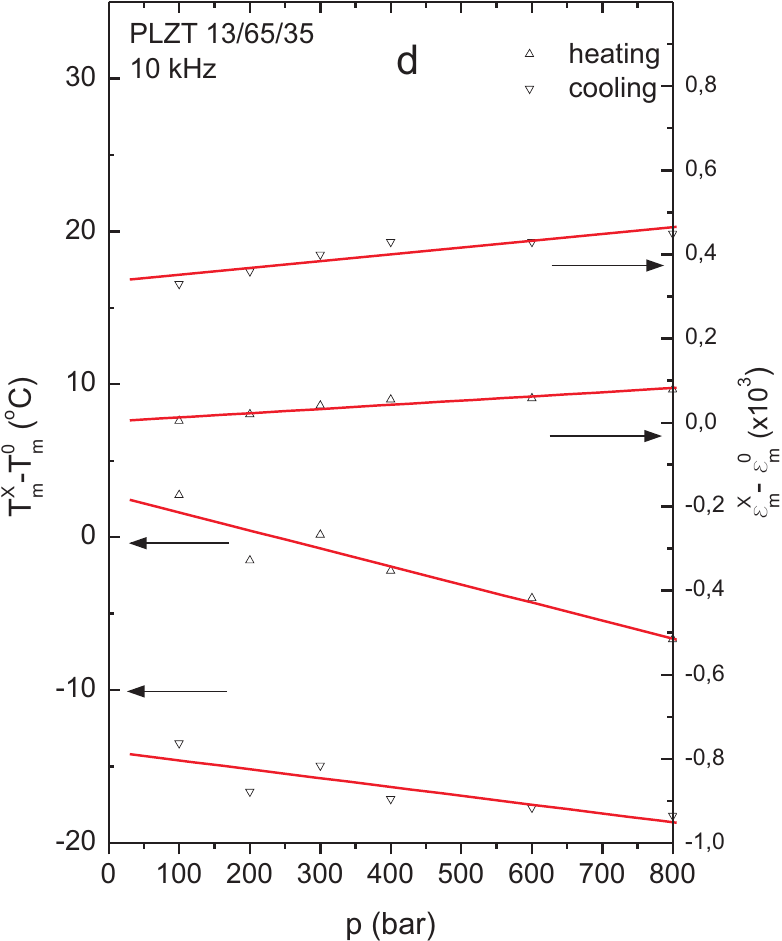}
}
\caption{Shift of the transition temperature and change of the maximal value of permittivity as a function of pressure of the   $T_\mathrm{m}^X-T_\mathrm{m}^0$ and   $\varepsilon_\mathrm{m}^X-\varepsilon_\mathrm{m}^0$ of PLZT-$x/65/35$ ceramics with $x=2$ (a), $6$ (b), $10$ (c) and $13$ (d) on heating and cooling. $T_\mathrm{m}^X$ and $T_\mathrm{m}^0$  is the transition temperature at pressure $X$ and $0$~bar, respectively. $\varepsilon_\mathrm{m}^X$  and $\varepsilon_\mathrm{m}^0$ is maximal value of the permittivity at the pressure $X$ and $0$~bar, respectively.}
\label{fig4}
\end{figure}

\begin{figure}[!tb]
\centerline{
\includegraphics[width=0.45\textwidth]{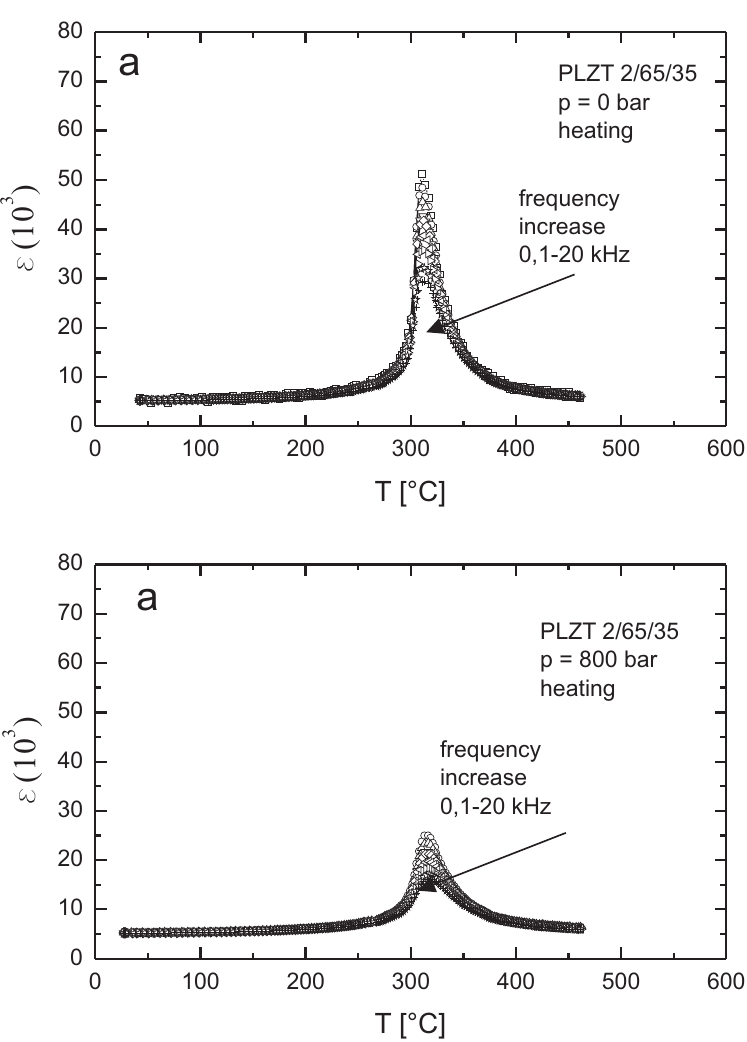}
\hspace{1mm}
\includegraphics[width=0.45\textwidth]{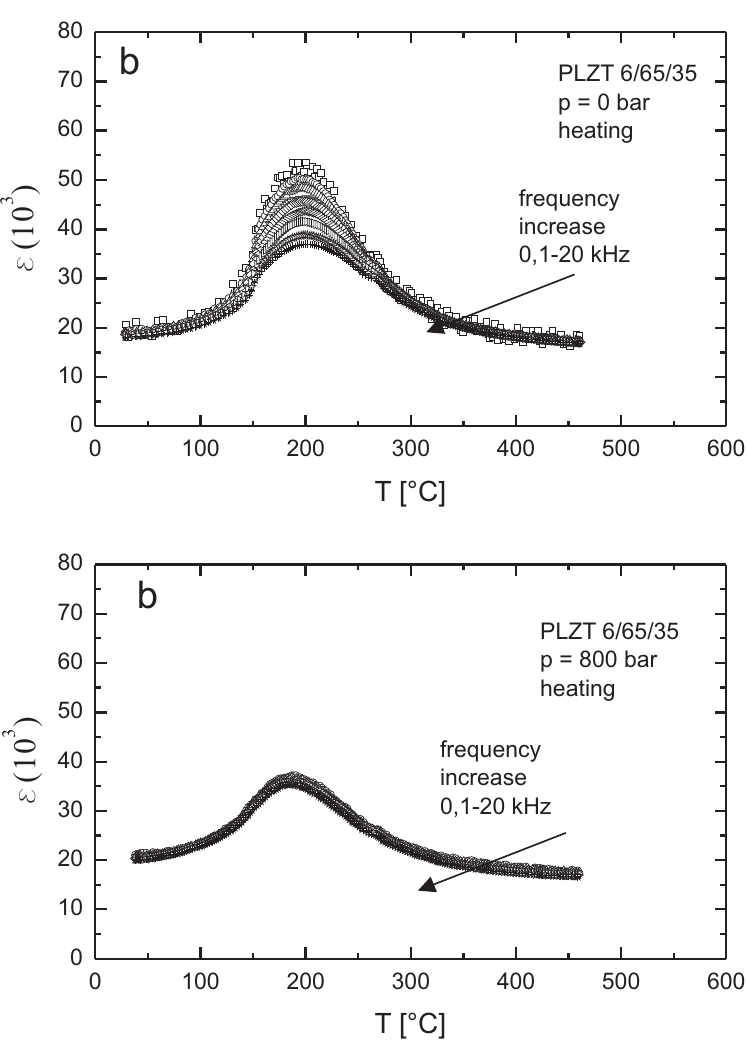}
}
\centerline{
\includegraphics[width=0.45\textwidth]{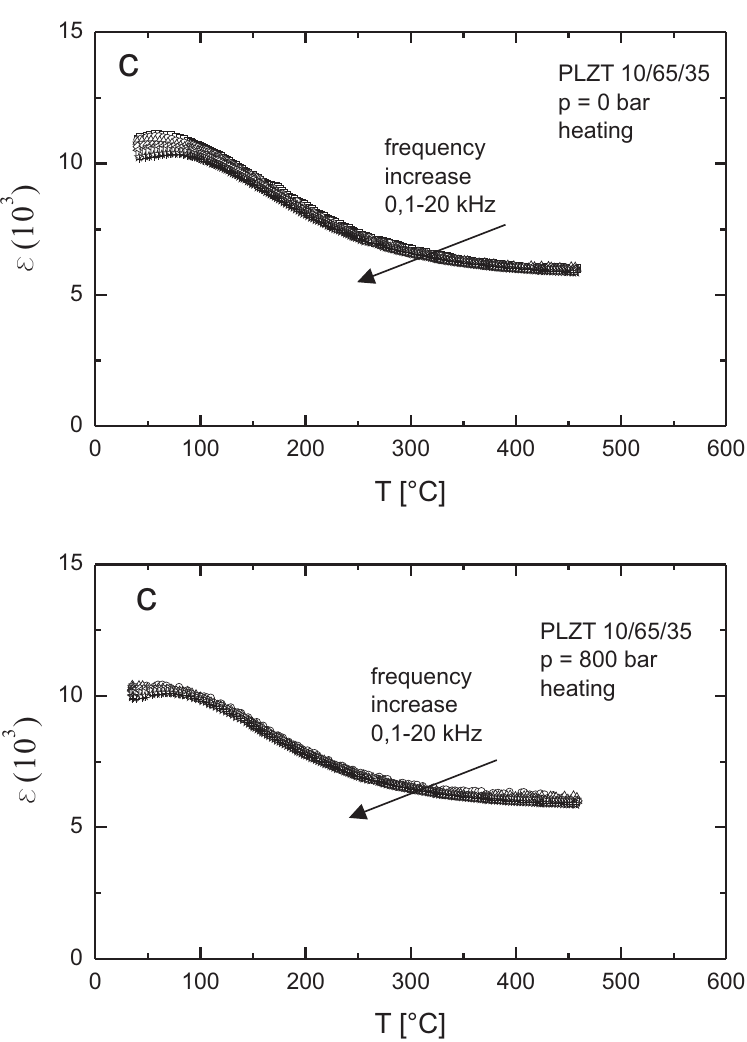}
\hspace{1mm}
\includegraphics[width=0.45\textwidth]{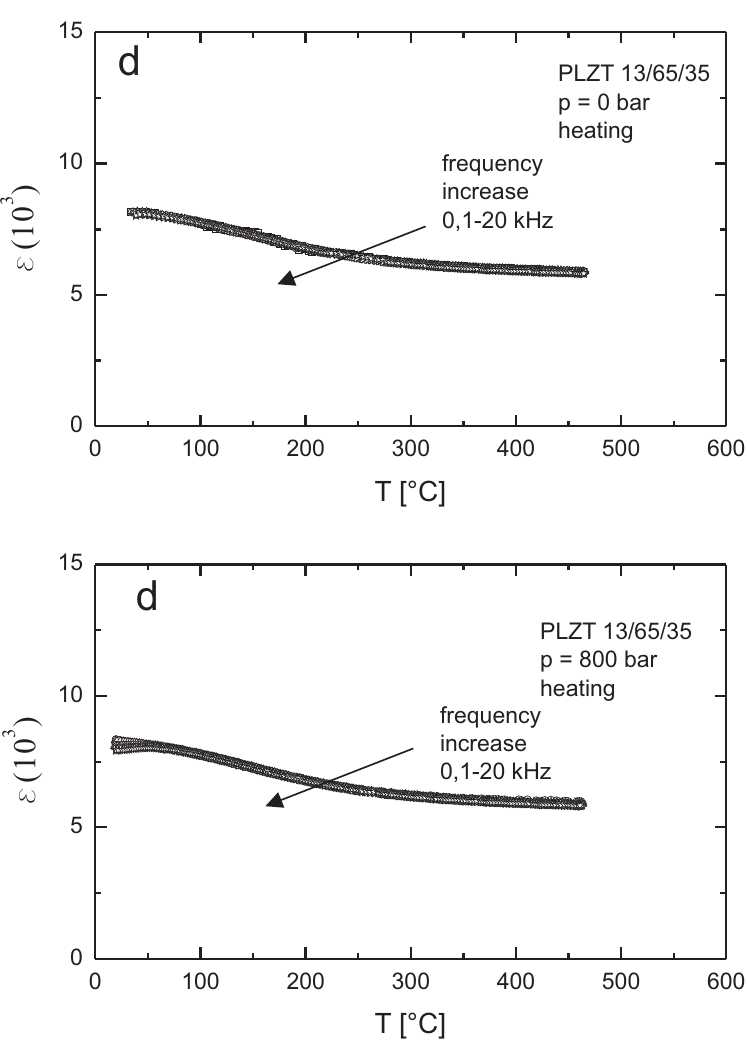}
}
\caption{Temperature/frequency dependence of electric permittivity for PLZT-$x/65/35$ ceramics ($x=2,\, 6,\, 10$ and $13$) on heating at $p=0$~bars (top) and $p=800$~bars (bottom).}
\label{fig5}
\end{figure}

The temperature/frequency dependence of the electric permittivity for the pressure applied parallel to the ac electric field for the PLZT ceramics analyzed are presented in figures~\ref{fig2}--\ref{fig5}. The main outcome obtained from  these dependences show that with an increase of pressure:
\begin{enumerate}
\item The $\varepsilon(T)$ maximum becomes more diffuse. The results in figure~\ref{fig2} reveal a remarkable reduction of the temperature of the maximum dielectric permittivity ($T_\mathrm{m}$) and an increase in the diffuseness of the dielectric permittivity peak by increasing the La$^{3+}$ content on the Pb$^{2+}$ site. A further result of the pressure application is the reduction of thermal hysteresis (inserts in figure~\ref{fig2}). These can suggest the character of the transformation change to the second order.

\item The temperature of $\varepsilon(T)$ maximum ($T_\mathrm{c}$) shifts toward higher temperature of approximately \linebreak $\partial T_\mathrm{c} / \partial p =5.2 \pm 0.5$\SI{}{\degreeCelsius}/kbar for PLZT-$x/65/35$ with $x=2$ and downward toward lower temperature of approximately $\partial T_\mathrm{c} / \partial p=-13.5 \pm 0.5$\SI{}{\degreeCelsius}/kbar, $\partial T_\mathrm{c} / \partial p=-8.6 \pm 0.5$\SI{}{\degreeCelsius}/kbar and $\partial T_\mathrm{c} / \partial p=-9.7 \pm 0.5$\SI{}{\degreeCelsius}/kbar for PLZT-$x/65/35$ with $x=6,\, 10$ and $13$, respectively, on heating.
The temperature of $\varepsilon(T)$ maximum ($T_\mathrm{c}$) shifts toward higher temperature of approximately $\partial T_\mathrm{c} / \partial p=7.4 \pm 0.5$\SI{}{\degreeCelsius}/kbar for PLZT-$x/65/35$ with $x=2$ and downward toward lower temperature
of approximately $\partial T_\mathrm{c} / \partial p=-19.5\pm 0.5$\SI{}{\degreeCelsius}/kbar, $\partial T_\mathrm{c} / \partial p=-9.8 \pm 0.5$\SI{}{\degreeCelsius}/kbar
and $\partial T_\mathrm{c} / \partial p= -12.7 \pm 0.5$\SI{}{\degreeCelsius}/kbar, for PLZT-$x/65/35$ with $x=6,\, 10$ and $13$, respectively, on cooling. The up and down shifts  are nearly linear in the range of pressure from 0 to 1kbar for the analyzed PLZT-$x/65/35$ ceramics ($x=2,\, 6,\, 10$ and $13$) (figure~\ref{fig4} and table~\ref{tab1}).

\item The maximum value of the electric permittivity decreases gradually with an increasing La$^{3+}$ content for PLZT-$x/65/35$ with $x=6,\, 10$ and $13$ (figure~\ref{fig4} and table~\ref{tab1}).

\item The maximum value of the electric permittivity decreases gradually with an increasing applied pressure for PLZT-$x/65/35$ with $x=2,\, 6$ and $10$, and increases gradually with an increasing applied pressure for PLZT-$x/65/35$ with $x=13$.
The maximum values of the electric permittivity change approximately
$\partial \varepsilon_\mathrm{m}/ \partial p= -16100 \pm 5$/kbar, $\partial \varepsilon_\mathrm{m}/ \partial p=-3400\pm 5$/kbar, $\partial \varepsilon_\mathrm{m}/ \partial p=-260 \pm 5$/kbar and $\partial \varepsilon_\mathrm{m}/ \partial p= -81\pm5$/kbar for PLZT-$x/65/35$ with $x=2,\, 6,\, 10$ and $13$, respectively, on heating.
The maximum value of the electric permittivity changes approximately $\partial \varepsilon_\mathrm{m}/ \partial p=-13000\pm5$/kbar, $\partial \varepsilon_\mathrm{m}/ \partial p=-3400\pm5$/kbar, $\partial \varepsilon_\mathrm{m}/ \partial p=-400\pm5$/kbar and $\partial \varepsilon_\mathrm{m}/ \partial p=120\pm5$/kbar for PLZT-$x/65/35$ with $x=2,\, 6,\, 10$ and $13$, respectively, on cooling.
The changes of maximal value of permittivity caused by the uniaxial pressure show that shifts are nearly linear in the range of pressure from 0 to 1~kbar for all the PLZT ceramics analyzed.

\item The dielectric dispersion decreases with an increasing La$^{3+}$ content for
PLZT-$x/65/35$ with $x=2,\, 6,\, 10$ and $13$.
The values of the electric dispersion change approximately from $\partial \varepsilon_\mathrm{m}/ \partial f=955$/kHz at $p=0$~bar to about 540/kHz at $p=1$~kbar, from $\partial \varepsilon_\mathrm{m}/ \partial f=830$/kHz at $p=0$~bar to about 90/kHz at $p=1$~kbar, from $\partial \varepsilon_\mathrm{m}/ \partial f=39$/kHz at $p=0$~bar to about 18/kHz at $p=1$~kbar, and from $\partial \varepsilon_\mathrm{m}/ \partial f= 0.01$/kHz at $p=0$~bar to about 8/kHz at $p=1$~kbar for PLZT-$x/65/35$ with $x=2,\, 6,\, 10$ and $13$, respectively.

\item The dielectric dispersion decreases with an increasing applied pressure for PLZT-$x/65/35$ with $x=2,\, 6$ and $10$.

\item Dielectric losses change in a way similar to dielectric constant. The maximum intensity of the $\tan\delta$ curve decreases, becomes more diffuse and shifts toward lower temperature for PLZT-$x/65/35$ with $x=2,\, 6,\, 10$ and $13$ (figure~\ref{fig3}).
\end{enumerate}

The samples, which are more sensitive to pressure are PLZT-$x/65/35$ with $x=6$ (the highest $\partial T_\mathrm{c} / \partial p$ on cooling) and PLZT-$x/65/35$ with $x=2$ (the highest $\partial \varepsilon_\mathrm{m}/ \partial p$ on heating). The sample, which is also more sensitive to the frequency of the measured electric field (the highest $\partial \varepsilon_\mathrm{m}/ \partial f$) is PLZT-$x/65/35$ with $x=2$ (figure~\ref{fig5} and table~\ref{tab1}).

\begin{table}[htb]
\caption{$\partial T_\mathrm{c} / \partial p$, $\partial \varepsilon_\mathrm{m}/ \partial p$ and $\partial \varepsilon_\mathrm{m}/ \partial f$ estimated for investigated materials.}
\label{tab1}
\begin{center}
\begin{tabular}{|r|r|r|r|}
\hline
Sample & $\partial T_\mathrm{c} / \partial p$ & $\partial \varepsilon_\mathrm{m}/ \partial p$ & $\partial \varepsilon_\mathrm{m}/ \partial f$ (heating)\\
 & (heating/cooling) & (heating/cooling) & $p=0$~bar/$p=1$~kbar \\ \hline\hline
PLZT  2/65/35 & $5.2/7.4\pm0.5$\SI{}{\degreeCelsius}/kbar & $-16100/-13000\pm5$/kbar & 955/540/kHz \\
PLZT  6/65/35 & $-13.5/-19.5\pm0.5$\SI{}{\degreeCelsius}/kbar & $-3400/-3400\pm5$/kbar & 830/90/kHz \\
PLZT 10/65/35 & $-8.6/-9.8\pm0.5$\SI{}{\degreeCelsius}/kbar & $-260/-400\pm5$/kbar & 39/18/kHz \\
PLZT 13/65/35 & $-9.7/-12.7\pm0.5$\SI{}{\degreeCelsius}/kbar & $81/120\pm5$/kbar & 0/8/kHz \\ \hline
\end{tabular}
\end{center}
\end{table}

The changes in properties of ceramics under the mechanical load ($0 \div 1000$~bars) can be attributed  to the creation or annihilation of detects, an elastic change of distances between ions in the crystal structure and to a change in the domain structure.
It is difficult to estimate the contribution of each mechanism. The contribution coming from the changes in the density of defects can be negligible because the changes of the properties are reversible after heating to high temperature followed by cooling \cite{1}.
Changes in the distance between ions can lead to the variations of phase transition temperature by changes of interaction constants or by changes of dipole moments. The phase transition temperature will increase or decrease depending on which mechanism is predominant \cite{6}.
Changes in the domain structure, i.e., induced domain wall movement and domain switching may be derived from applying a mechanical stress. Mechanical load in the investigated materials is large enough to reduce the density of the domains in the direction parallel to the stress by non-\SI{180}{\degree} domain switching and to increase the density of the domain in perpendicular direction. This implies a decrease of the domain ordering in one direction and an increase of the domain ordering in the opposite direction. Electric and elastic nanoregions in PLZT can be reoriented by uniaxial pressure in the direction in which pressure is applied for PLZT-$x/65/35$ with $x>$6. Nanoregions are forced in the new positions and their contribution to the electric permittivities is smaller, and thus $\varepsilon$ and tan$\delta$ decrease (figures~\ref{fig2}--\ref{fig5}). A further effect of the applied pressure can be an increase in sizes of nanoregions and their combination into larger complexes, which results in a weaker dielectric response. These effects can cause a narrowing distribution of relaxation time, which in turn can lead to a suppression of  dispersion in the frequency range used.

For a normal ferroelectric above the temperature of the phase transition ($T_\mathrm{c}$), the electric permittivity falls off with temperature according to:
\begin{equation}
\varepsilon=\varepsilon_{0}+\frac{C}{T-T_{0}}\simeq\frac{C}{T-T_{0}}\,,
\end{equation}
where $C$ is the Curie constant and $T_{0}$ is the Curie-Weiss temperature. Both $C$ and $T_{0}$ decrease with an increasing pressure. It was found that for our samples ($p=0$ bar), $\varepsilon$ starts to obey the Curie-Weiss law at temperatures higher than $T_\mathrm{m}$ (e.g. temperature at which $\varepsilon$ reaches the maximum). Deviation degree of $\varepsilon$ from the Curie-Weiss law can be defined by $\Delta T_\mathrm{cm}$ as follows: $\Delta T_\mathrm{cm}=T_\mathrm{dev}-T_\mathrm{m}$, where $T_\mathrm{dev}$ is the temperature, at which $\varepsilon$ starts to obey the Curie-Weiss law. It is found that for $p=0$ bar, $\Delta T_\mathrm{cm} =5.5$\SI{}{\degreeCelsius}, 14\SI{}{\degreeCelsius}, 18\SI{}{\degreeCelsius}, and 23\SI{}{\degreeCelsius} for PLZT-$x/65/35$ with $x=2,\, 8,\, 10$ and $13$, respectively. For $p=800$~bar: $\Delta T_\mathrm{cm} =2.5$\SI{}{\degreeCelsius}, 13\SI{}{\degreeCelsius}, 26\SI{}{\degreeCelsius}, and 32\SI{}{\degreeCelsius} for PLZT-$x/65/35$ with $x=2,\, 8,\, 10$ and $13$, respectively. This implies that the applied uniaxial pressure weakens (for PLZT-$x/65/35$ with $x=2$ and $8$) or enhances (for PLZT-$x/65/35$ with $x=10$ and $13$) the diffuse phase transformation behavior.

A modified expression of Curie-Weiss law explains the diffuseness of ferroelectric phase transition described by:
\begin{equation}
\label{eq2}
\frac{1}{\varepsilon}-\frac{1}{\varepsilon_\mathrm{m}}=\frac{(T-T_\mathrm{m})^{\gamma}}{C^{*}}\,,
\end{equation}
where $\gamma$ and $C^{*}$ are constants, $\gamma$ value is between 1 and 2. The limiting value $\gamma$=1 reduces the equation to normal Curie-Weiss law (for classical ferroelectrics) and $\gamma=2$ reduces the equation to quadratic (for relaxor ferroelectrics) \cite{7}. The dependence of $\ln(1/\varepsilon-1/\varepsilon_\mathrm{m})$ on $\ln(T-T_\mathrm{m})$ according to equation~(\ref{eq2}) for the analyzed ceramics shows the slope of the curve giving the diffuseness constant $\gamma$. A nearly linear correlation is observed on heating for $f=10$~kHz and for $p=0$~bar there is obtained $\gamma=1.7$, $\gamma=1.1$, $\gamma=1.7$ and, $\gamma=1.2$ for PLZT-$x/65/35$, $x=2,\, 6,\, 10$ and $13$, respectively. However, for higher pressures, $\gamma$ increases indicating that a classical ferroelectric seems to change to a relaxor one. For $p=800$~bar there is obtained $\gamma=1.9$, $\gamma=1.7$, $\gamma=1.9$, $\gamma=1.9$ for PLZT-$x/65/35$, $x=2,\, 6,\, 10$ and $13$, respectively.

PLZT ceramics have perovskite-type ABO$_{3}$ structure with Zr/Ti ions located at the place of the B cations and Pb/La ions occupying the A-sites. Pb ions displace in the direction of smaller Ti ions. The positive charge of the Ti ion, whose atomic radius is smaller than that of Zr ion, is better screened by surrounding oxygen ions and allows a shorter Pb-Ti distance \cite{8}. The incorporation of Ti ions instead of Zr to PbZrO$_{3}$ leads to an increase of the phase transition temperature in PZT system. Due to a smaller ionic radius of the Ti ion than Zr one, the material was subjected to compressive stress with an increased Ti content. This could explain why the compressive stress induces a similar effect as that from an increase of the concentrations of Ti ions in PZT.

\section{Conclusions}

High density PLZT-$x/65/35$ ($x=2,\, 6,\, 10$ and $13$) ceramics with the average grain size of ca. $5\div8$~\SI{}{\micro\metre} was obtained by a two-stage hot-pressing technology. The effect of the uniaxial pressure on dielectric response of these ceramics has been examined. There is a diffuse transition in the examined PLZT ceramics, and the diffuseness constant $\gamma$ depends on pressure. With an increase of pressure, the phase transition temperature, the frequency dispersion and the thermal hysteresis change, and the dielectric behavior characteristics spread. Applying uniaxial pressure or increasing the Ti content in a PZT system would induce similar effects.

\newpage
\ukrainianpart

\title{Діелектричні властивості  PLZT-$x/65/35$ ($2\leqslant x \leqslant
13$) під впливом механічного напруження, електричного поля і
температурного навантаження}

\author{К.~Питель\refaddr{ad1}, Я. Суханич\refaddr{ad2}, М. Лівінш\refaddr{ad3}, А. Стернберг\refaddr{ad3}}

\addresses{
\addr{ad1}Інститут технології, Педагогічний університет,  30--084
Краків, Польща %
\addr{ad2} Інститут фізики, Педагогічний університет,
30--084 Краків, Польща %
\addr{ad3} Інститут фізики твердого тіла,
Університет Латвії, 1063 Рига, Латвія}

\makeukrtitle
\begin{abstract}

Ми дослідили вплив одновісного тиску  ($0 \div 1000$~bars),
прикладеного паралельно  до  змінного електричного поля, на властивості
кераміки PLZT-$x/65/35$ ($2 \leqslant x \leqslant 13$). Виявлено
значний вплив зовнішнього напруження на ці властивості. Прикладання
одновісного тиску веде до зміни піку інтенсивності електричної
проникності  ($\varepsilon$), частотної дисперсії і діелектричного
гістерезису. Пік інтенсивності  $\varepsilon$ стає
розмитий/загострений і зсувається до вищих/нижчих температур з ростом
тиску. Робиться висновок, що прикладання одновісного тиску індукує
подібні ефекти як підвищення концентрації іонів  Ti в системі PZT.
Ми інтерпретуємо наші результати на основі процесів  перемикання
доменів під дією комбінованого електромеханічного навантаження.
\keywords сегнетоелектрик, PLZT-x/65/35, діелектричні властивості, одновісний тиск
\end{abstract}

\end{document}